\documentclass[]{aa}
\usepackage{natbib}
\usepackage{graphicx}
\usepackage[varg]{txfonts}
\begin{document}

   \title{Modelling variability of solar activity cycles}
   \author{L. L. Kitchatinov\inst{1,2}
          \and
          A. V. Mordvinov\inst{1}
          \and
          A. A. Nepomnyashchikh\inst{1}
          }

   \institute{Institute for Solar-Terrestrial Physics, Lermontov Str. 126A, Irkutsk, 664033, Russia\\
              \email{kit@iszf.irk.ru}
         \and
             Pulkovo Astronomical Observatory, St. Petersburg, 196140, Russia
             }

   \date{Received 26 December 2017; accepted 27 March 2018}
  \abstract
{Solar activity cycles vary in amplitude and duration. The variations can be at least partly explained by fluctuations in dynamo parameters.} 
{We want to restrict uncertainty in fluctuating dynamo parameters and find out which properties of the fluctuations control the amplitudes of the magnetic field and energy in variable dynamo cycles.} 
{A flux-transport model for the solar dynamo with fluctuations of the Babcock-Leighton type $\alpha$-effect was applied to generate statistics of magnetic cycles for our purposes. The statistics were compared with data on solar cycle periods to restrict the correlation time of dynamo fluctuations.} 
{A characteristic time of fluctuations in the $\alpha$-effect is estimated to be close to the solar rotation period. The fluctuations produce asymmetry between the times of rise and descent of dynamo cycles, the rise time being on average shorter. The affect of the fluctuations on cycle amplitudes depends on the phase of the cycle in which the fluctuations occur. Negative fluctuations (decrease in $\alpha$) in the rise phase delay decay of poloidal field and increase the cycle amplitude in toroidal field and magnetic energy.
Negative fluctuation in the decline phase reduces the polar field at the end of a cycle and the amplitude of the next cycle. The low amplitude of the 24th solar cycle compared to the preceding 23rd cycle can be explained by this effect. Positive fluctuations in the descent phase enhance the magnetic energy of the next cycle by increasing the seed poloidal field for the next cycle. The statistics of the computed energies of the cycles suggest that superflares of $\ge 10^{34}$\,erg are not possible on the Sun.} 
{} 

   \keywords{Sun: dynamo -- Sun: activity -- Sun: magnetic fields}

   \maketitle
\section{Introduction}
Cycles of solar magnetic activity are known to vary in strength and duration. The variations are most probably attributed to the intrinsic randomness of the solar dynamo mechanism \citep[][Sect.\,5]{H88,C10}. Of the two basic dynamo effects of toroidal field winding by differential rotation and conversion of the toroidal field back to a poloidal field by cyclonic motions, the latter  is believed to be random to a large extent \citep{H93}.

There is extensive literature on dynamos with fluctuating parameters including recent publications by \citet{USM09}, \citet{CK12}, \citet{OK13}, \citet{Pea14}, \citet{CS17}, and \citet{IAR17}. The publications are mainly focussed on durable epochs of exceptionally low or high activity known as grand solar minima or maxima \citep{USK07}. This paper concerns another subject. The memory time of the dynamo process is approximately the duration of the activity cycle \citep{KN12}. It is therefore important to identify those properties of a fluctuating dynamo that are responsible for the increases or decreases in strengths between neighbouring cycles. This paper uses a flux transport dynamo model with fluctuations in the Babcock-Leighton (BL) mechanism of poloidal field generation for this aim.

Various dynamo-related parameters such as turbulent diffusivity or velocity of the  meridional flow can fluctuate and contribute to the variability of activity cycles \citep{CK12}. The primary contribution is, however, expected to come from fluctuations in the BL mechanism because of their relatively large amplitude. The BL mechanism is related to finite average tilts of solar active regions relative to the lines of latitude \citep{B61}. Its large fluctuations are caused by two properties of solar active regions:  the broad distribution of their tilt angles \citep[cf. fig.\,11 in][]{H96} and the moderate number of active regions simultaneously present on the Sun. The mechanism is, therefore, driven by a small ensemble of random objects.

Computations of \citet{Nea17} have suggested that even a single \lq\lq rogue'' active region can influence strength of the subsequent solar cycles. \citet{JCS15} pointed to AR10696 as a characteristic example of the active regions with abnormal tilts that are responsible for the weakness of solar cycle 24.

The amplitude of fluctuations in the BL mechanism can be estimated from sunspot data \citep{OCK13,JCS14}. The characteristic duration of the fluctuations is, however, less certain. We constrain the correlation time by comparing the statistics of the computed and observed durations of solar activity cycles. The dynamo model then includes random variations with time in the BL-type $\alpha$-effect with the so-defined amplitude and characteristic duration. The model computations show that the effect of fluctuations depends not only on their sense (increase or decrease) but also on the phase of a dynamo cycle in which the fluctuations occur. The phase dependence is caused by the difference in the dynamo process between ascending and descending phases of the activity cycles and can be interpreted in terms of the basic dynamo mechanisms. The interpretation generally agrees with the observed dynamics of sunspot activity and polar magnetic fields of recent activity cycles.

The next section describes our dynamo model and the method of allowance for random fluctuations. Section \ref{RD} presents and discusses the results of the modelling. This section also compares some of our findings with observations. The final Sect.\,\ref{Conclusions} summarises our conclusions.
\section{Dynamo model}
Our dynamo model belongs to the so-called flux-transport models initiated by \citet{WSN91} and first developed by \citet{CSD95} and \citet{D95}. This name reflects the importance of meridional flow for latitudinal transport of magnetic fields in this class of dynamo model. Flux-transport models are consistent with basic solar observations \citep{Jea13}.

Apart from an allowance for fluctuations in the $\alpha$-effect, the model of this paper is identical to that of \citet{KN17MNRAS,KN17AL}. We therefore describe in detail only the method of simulation of the random temporal variations in the $\alpha$-effect but do not repeat the model equations, all of which are given and commented on in the above-quoted papers.
\subsection{Model design}
The 2D dynamo model borrows the differential rotation and meridional flow from the differential rotation model of \citet{KO11}. The differential rotation is very close to seismological inversions for internal solar rotation. The one-cell meridional flow is also close to the seismological detection by \citet{RA15}, which is distinct among other recent detections in that this detection satisfies the basic condition of mass conservation.

The eddy diffusivity profile is derived from the entropy gradient, which is a  side product of the differential rotation model. The diffusivity varies smoothly around the value of $3\times 10^{12}$\,cm$^2$s$^{-1}$ in the bulk of the convection zone. Observation-based estimations by \citet{CS16} give a similar value. The magnetic diffusivity  however drops by almost four orders of magnitude in a thin ($\simeq 0.025R_\sun$) layer  near to the bottom boundary. The diffusivity drop near to the bottom of the convection zone is important for the diamagnetic pumping effect, which is significant for the performance of solar dynamo models \citep{KKT06,GG08}. Downward diamagnetic pumping of our model concentrates the toroidal field at the near-bottom layer of low diffusion. This reduces the rate of the field diffusive decay and provides a combination of relatively weak ($\sim 10$\,G) polar fields with about three orders of magnitude stronger toroidal fields near to the bottom of the convection zone \citep{KO12}. The eddy diffusion and diamagnetic pumping are anisotropic in our model. The anisotropy is induced by rotation. This means that the diffusivity along the rotation axis is somewhat larger than the diffusivity for the direction normal to the axis, and the diamagnetic pumping rate and direction depend on the magnetic field orientation relative to the rotation axis \citep{Oea02}.

The $\alpha$-effect of our model is non-local in space: the poloidal field near to the surface is generated from the bottom toroidal field. The non-local effect is supposed to represent the BL mechanism. The parameter $\alpha$ of the $\alpha$-effect includes the algebraic magnetic quenching\begin{equation}
    \alpha = \frac{\alpha_0}{1 + (B(r_\mathrm{i},\theta)/B_0)^2} f(r,\theta) ,
    \label{1}
\end{equation}
i.e. it depends on toroidal field $B$ and decreases with the field strength. The decrease can be thought of as resulting from a faster buoyant rise of the stronger field that reduces the effect of the Coriolis force. In Eq.\,(\ref{1}), $r$ is the radius, $\theta$ is the co-latitude, the function $f(r,\theta)$ is specified in \citet{KN17MNRAS}, and the parameter $B_0 = 10$\,kG in all our computations. Algebraic quenching of Eq.\,(\ref{1}) is the only non-linearity in our model. The back reaction of the generated magnetic fields on the differential rotation or meridional flow is neglected.

The boundary conditions correspond to the interface with a superconductor at the bottom and to the vertical field (zero toroidal, radial poloidal) at the top.

The critical value of the $\alpha_0$-parameter of Eq.\,(\ref{1}) in our model is $\alpha_0^\mathrm{cr} = 0.158$\,m\,s$^{-1}$. Non-decaying magnetic cycles are found only for $\alpha_0$ exceeding this value. Observations of stellar rotation show that the solar rotation rate is not far above the threshold rate for global dynamos \citep{MvS17}. Estimations based on observational data on stellar rotation suggest that the solar dynamo is about 10\% supercritical \citep{KN17MNRAS}. Therefore, $\alpha_0 = 0.174$\,m\,s$^{-1}$ in our computations.

An initial field of mixed parity was prescribed. The field dynamics  forgets the initial condition after several diffusion times and the field approaches dipolar (equator-antisymmetric) parity. All results in  Sect.\,\ref{RD} correspond to such an asymptotic regime.
\subsection{Fluctuating $\alpha$-effect}
To allow for random variations in the $\alpha$-effect with time, parameter $\alpha$ of Eq.\,(\ref{1}) is changed as follows:
\begin{equation}
    \alpha \longrightarrow \alpha\ \left( 1\ +\ \sigma s(t)\right) ,
    \label{2}
\end{equation}
where $\sigma$ is the relative amplitude of fluctuations and $s(t)$ is the random function of time of order one.

Following \citet{R05}, we simulate this random function by solving numerically the equation system
\begin{eqnarray}
    \frac{\mathrm{d}s}{\mathrm{d}t} &=& -\frac{n}{\tau}\left( s - s_1\right)\ ,
    \nonumber \\
    \frac{\mathrm{d}s_1}{\mathrm{d}t} &=& -\frac{n}{\tau}\left( s_1 - s_2\right)\ ,
    \nonumber \\
     ... &&
    \nonumber \\
    \frac{\mathrm{d}s_{n-1}}{\mathrm{d}t} &=& -\frac{n}{\tau}\left( s_{n-1} - \sqrt{\frac{2\tau}{\Delta t}}\ \hat{g}\right)\ ,
    \label{3}
\end{eqnarray}
in line with the dynamo equations. In Eqs.\,({\ref{3}), $\tau$ is the correlation time, $\Delta t$ is the numerical time-step, and $\hat{g}$ is the normally distributed random number with zero mean and $rms$ value equal one. The value of $\hat{g}$ is renovated on each time step independently of its previous value. The random function $s(t)$ is therefore initiated by a short-correlated ($\Delta t \ll \tau$) random forcing. This function varies continuously with time, although its $n$-th order time derivative is discontinuous. For $n$ equal 1 or 2 and short time-step $\Delta t \ll \tau$ (that is the case with our computations), the correlation function $\phi(t) = \langle s(t_0 + t)s(t_0)\rangle$ can be derived analytically \citep{OK13}; analytical results are confirmed numerically. The factor in front of $\hat{g}$ in the last of the Eqs.\,(\ref{3}) is chosen so that $\phi(0) = \langle s^2(t) \rangle = 1$ for the analytical correlation functions.

Correlation time $\tau$ is the model parameter. Computations of this paper were performed with $n = 3$ in Eqs.\,(\ref{3}).
\subsection{What can be expected?}\label{S2.3}
The consequences of the two basic effects of $\alpha\Omega$-dynamos differ between the growth and decay phases of an activity cycle. In the growth phase, the toroidal field and magnetic energy are amplified by the $\Omega$-effect of the toroidal field winding from the pre-existing poloidal field by differential rotation. The amplification is restricted by the $\alpha$-effect, which reduces the poloidal field and causes it to reverse near to the activity maximum. It may be expected, therefore, that a reduced $\alpha$-effect would lead to a stronger cycle and an increased $\alpha$ in the growth phase would reduce the cycle amplitude. In the decay phase, the $\alpha$-effect generates the  seed poloidal field for the next cycle. It may be expected that a reduced $\alpha$-effect would result in a smaller poloidal field and a weaker next cycle. The opposite is expected from the fluctuation increasing the $\alpha$-effect on the descending activity phase. To test the expectations, we computed the field dynamics separately for ascending and descending phases of the magnetic cycles, artificially varying the $\alpha$-effect.

\begin{figure}
   \centering
   \includegraphics[width=7.7 truecm]{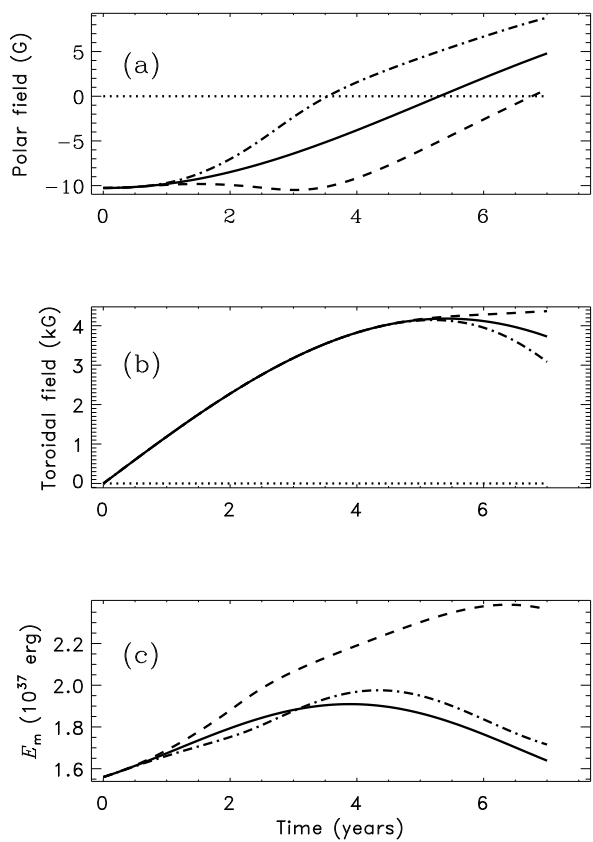}
   \caption{Poloidal field at the northern pole (a), bottom toroidal field
            $B_\mathrm{t}$ at $15\degr$ latitude (b), and magnetic energy (c) in ascending phase of a magnetic cycle computed with perturbation in the $\alpha$-effect imposed in a range of time between 0 years and 2 years. The full line shows the reference case of not varied $\alpha$, the dashed line indicates the reversed sign of $\alpha$, and the dash-dotted line shows the $\alpha$ increased by a factor of 3.
              }
   \label{f1}
\end{figure}

\begin{figure}
   \centering
   \includegraphics[width=7.7 truecm]{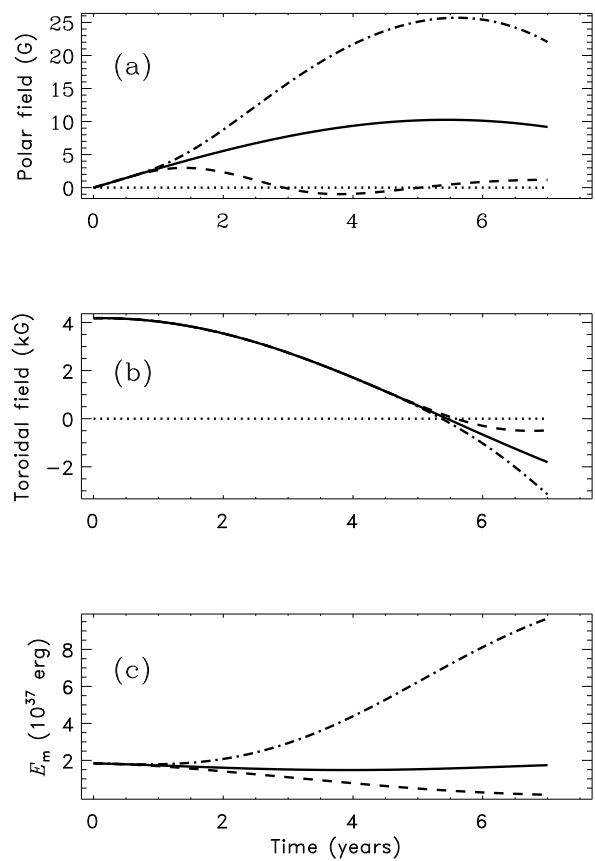}
   \caption{Same as in Fig.\ref{f1} but for descending phase of a cycle.
              }
   \label{f2}
\end{figure}

The strength of computed magnetic cycles is characterised by the bottom toroidal field $B_\mathrm{t}$ at latitude $15\degr$, close to the latitude where the bottom field attains its maximum strength. Another relevant parameter is the total energy of the toroidal field, that is
\begin{equation}
    E_\mathrm{m} = \frac{1}{4}\int\limits_{r_\mathrm{i}}^{r_\mathrm{e}}\int\limits_{-1}^{1}\ B^2(r,\cos\theta)r^2\ \mathrm{d}\cos\theta\ \mathrm{d}r\ ,
    \label{4}
\end{equation}
where $r_\mathrm{e} = 0.97R_\sun$ is the external boundary of the computation domain. The beginning of a new activity cycle is defined as the instant of sign reversal of $B_\mathrm{t}$. The cycle amplitude is defined as the maximum absolute value of $B_\mathrm{t}$ in a cycle.

To estimate expectations for the consequences of fluctuations in the $\alpha$-effect in the ascending phase of an activity cycle, the $\alpha_0$ value was changed abruptly at the beginning of the computed cycle and then returned to the initial value of $0.174$\,m\,s$^{-1}$ two years after the change. The effect of negative fluctuations was probed by reversing the sign of $\alpha_0$, i.e. by changing its value to $-0.174$\,m\,s$^{-1}$; the effect of positive fluctuations was probed by changing $\alpha_0$ in the opposite direction by the same amount, i.e. to $\alpha_0 = 0.522$\,m\,s$^{-1}$.

The resulting variations in the poloidal and toroidal fields and in magnetic energy are shown in Fig.\,\ref{f1}. The plots generally confirm the above expectations but with a reservation that the field reacts on variation in $\alpha$ with a delay. Polar field and magnetic energy react with the delay of about one year. The bottom field $B^\mathrm{t}$ needs about four years more to respond to the changes in the poloidal field generation. The slower reaction of $B^\mathrm{t}$ is probably explained by additional time required for the field transport to the bottom by the meridional flow and diffusion.

Figure\,\ref{f1} shows that the poloidal field reduces faster and reverses earlier with positive $\alpha$-fluctuation. The maximum value of $B^\mathrm{t}$ is almost unaffected by the positive fluctuation however. The effect in magnetic energy is small as well. The energy is reduced initially by the positive fluctuation but eventually increases slightly compared to the case of unchanged $\alpha$.

The negative fluctuation, on the contrary, supports the pre-existing poloidal field and delays its polar reversal. Accordingly, a stronger toroidal field is wound by the differential rotation and larger magnetic energy is produced. The toroidal field keeps on growing at the end of the run in Fig.\,\ref{f1}b implying a higher cycle amplitude in $B^\mathrm{t}$ compared with the case of unchanged $\alpha$.

The effect of fluctuations in the descending phase of a cycle depends strongly on their sense. This is illustrated by Fig.\,\ref{f2}. Positive fluctuation increases the polar field at the end of the magnetic cycle thus preparing a strong next cycle. It also increases the magnetic energy. Negative fluctuation acts in the opposite direction, producing a weak polar field at the end of the cycle and reducing the magnetic energy. Thus, the cycle following a descending phase dominated by negative $\alpha$-fluctuations is expected to be weak.

Regularities seen in Figs. \ref{f1} and \ref{f2} help to explain the results of computations with randomly varying $\alpha$ in the next section.
\section{Results and discussion}\label{RD}
As explained in the Introduction, fluctuations in the BL mechanism are  large. The amplitude of the fluctuations can be estimated from sunspot data. Estimations by \citet{OCK13} give the value of $\sigma = 2.7$ for the relative amplitude $\sigma$ of Eq.\,(\ref{2}). This value is used in our computations.

The correlation time $\tau$ of the fluctuations is less certain. The correlation time defines the statistical mean duration of the fluctuations. In the case of the BL mechanism, the correlation time can be thought of as the characteristic lifetime of the solar active regions. It is expected to be comparable with the period of solar rotation $P_\mathrm{rot} = 25.4$ days. We selected the appropriate correlation time by computing statistics of magnetic cycles for various $\tau/P_\mathrm{rot}$ ratios and comparing these statistics with statistics of the observed activity cycles. A comparison with the amplitudes of the observed cycles is problematic because a conversion of field strength or energy of the dynamo model into the observed sunspot numbers or areas is uncertain. A comparison with the observed cycle periods seems to be straightforward.
\subsection{Cycle periods}\label{S3.1}
The statistics of 10\,000 magnetic cycles were computed with $\tau/P_\mathrm{rot} = 0.5, 1, 1.5$ and $2$. The variability coefficient for the cycle period
\begin{equation}
    D = \frac{1}{\langle P_\mathrm{cyc}\rangle}\langle\left(P_\mathrm{cyc}
    - \langle P_\mathrm{cyc}\rangle\right)^2\rangle^{1/2}
    \label{5}
\end{equation}
varied as $0.104,\ 0.162,\ 0.212,$ and $0.248$, respectively.
The angular brackets in this equation signify averaging over the ensemble of 10\,000 computed cycles. More durable fluctuations naturally produce a larger variability.

Periods of 36 cycles only are known from direct observations (https://www.ngdc.noaa.gov/stp/space-weather/solar-data/solar-indices/sunspot-numbers/cycle-data/table\_cycle-dates\_maximum-minimum.txt).
The variability of Eq.\,(\ref{5}) $D_\mathrm{obs} = 0.135$ for this sample. Periods for a larger number of 119 cycles (measured maximum-to-maximum) have been recovered by \citet{Nea15} from solar activity proxies; $D_\mathrm{rec} = 0.185$ for their sample. These values are close to $D = 0.162$ resulting from our model computations for $\tau = P_\mathrm{rot}$. Unless otherwise stated, the results to follow refer to this case of the correlation time equal to the rotation period.

\begin{figure}
   \includegraphics[width=\hsize]{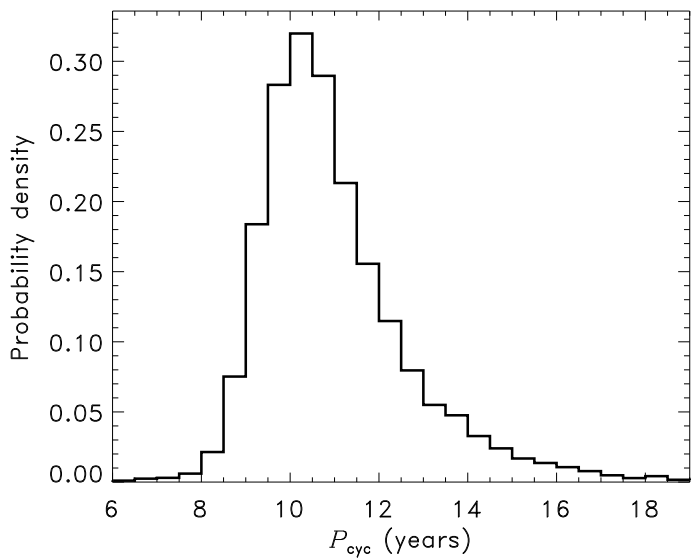}
   \caption{Distribution function for durations of computed cycles for the case of $\tau = P_\mathrm{rot}$.
              }
   \label{f3}
\end{figure}

Figure\,\ref{f3} shows the distribution function for the computed cycle periods. The distribution is similar to the observations-based results of \citet{Nea15}. The mean duration of computed cycles $\langle P_\mathrm{cyc}\rangle = 10.9$ years.

It is remarkable that the mean duration of the growth phase $\langle P_\mathrm{gr}\rangle = 5.1$ years is shorter than the decay phase $\langle P_\mathrm{dec}\rangle = 5.8$ years, although both phases were almost equally durable ($\simeq 5.4$ years) in the model without fluctuations. This asymmetry was found in computations with all tried values of the correlation time $\tau$ and it increases steadily with $\tau$. The effect of fluctuations of the BL mechanism depends on the phase of the dynamo cycle in which the fluctuations occur. It is therefore not surprising that the fluctuations induce asymmetry between growth and decay phases. Similar to the observed solar cycles, the asymmetry is a statistical effect: not all computed cycles have relatively short growth pases but they are shorter than the decay phases on average.
The Waldmeier effect of the inverse correlation between cycle amplitude and its rise time is not reproduced with our model.
\subsection{Regularities of irregular cycles}
The model without fluctuations gives a sequence of exactly repeatable cycles. The cycles have the amplitude of $B_\mathrm{t} = 4.18$\,kG in the toroidal field and $E_\mathrm{m} = 1.91\times10^{37}$\,erg in magnetic energy. The polar field reaches its maximum amplitude of $B_\mathrm{p} = 10.3$\,G close to the instant of $B_\mathrm{t}$ reversal. The toroidal field amplitude is therefore related to the polar field of the preceding minimum by the equation
\begin{equation}
    B_\mathrm{t} = -407 B_\mathrm{p}\ ,
    \label{6}
\end{equation}
which accounts for the opposite sign of $B_\mathrm{p}$ and $B_\mathrm{t}$ in the same hemisphere. A similar relation between magnetic energy and polar field is written
\begin{equation}
    E_\mathrm{m} = 1.8\times10^{35} B_\mathrm{p}^2 .
    \label{7}
\end{equation}
The model with fluctuations produces cycles of variable amplitude. The variable cycles follow closely the relation (\ref{6}) (Fig.\,\ref{f4}). The correlation between the cycle strengths and poloidal field of the preceding minima is well known from solar observations \citep{MT00,Sea05} and explained in dynamo theory \citep{Sea78,CCJ07}.

\begin{figure}
   \includegraphics[width=\hsize]{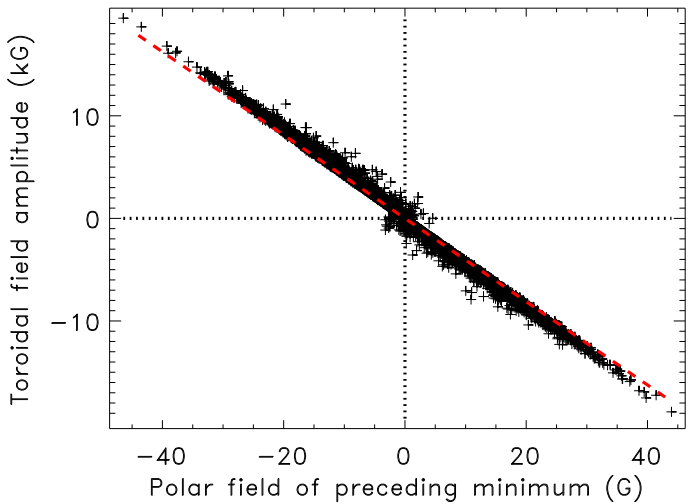}
   \caption{Scatter plot of computed magnetic cycles on the plane of the polar
            field of a cycle minimum and the amplitude of $B_\mathrm{t}$ in the next cycle. Relation of Eq.\,(\ref{6}) is overplotted by the dashed line.
              }
   \label{f4}
\end{figure}

Except for weak cycles in the near-centre region of Fig.\,\ref{f4} (Grand minima), the relation (\ref{6}) represents the low bound for the absolute value of $\mid B_\mathrm{t}/B_\mathrm{p}\mid$ ratio.

\begin{figure}
   \includegraphics[width=\hsize]{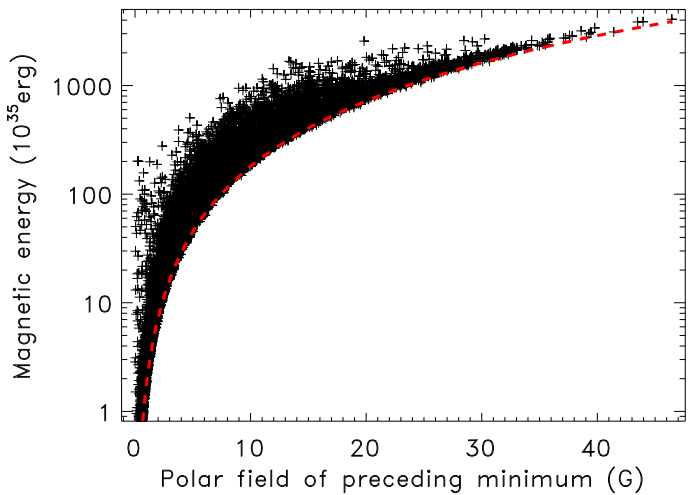}
   \caption{Scatter plot of computed magnetic cycles on the plane of the modulus of the polar
            field of a cycle minimum and the maximum magnetic energy (\ref{4}) in the next cycle. Relation of Eq.\,(\ref{7}) is overplotted by the dashed line.
              }
   \label{f5}
\end{figure}

Figure\,\ref{f5} shows that Eq.\,(\ref{7}) gives the low bound for the amplitude of cycle energy as a function of the polar field of the preceding minimum. This agrees with expectations of Sect.\,\ref{S2.3}: fluctuations on the ascending phase of a cycle increase the magnetic energy (Fig.\,\ref{f1}). Figure\,\ref{f5} shows larger scatter and larger deviations from the dashed line compared to Fig.\,\ref{f4}. The reason probably is that the magnetic energy (\ref{4}) depends not only on the field magnitude but on its spatial distribution as well. Stronger solar cycles are known to show broader distributions of sunspots \citep{Mea17}.

\begin{figure}
   \includegraphics[width=\hsize]{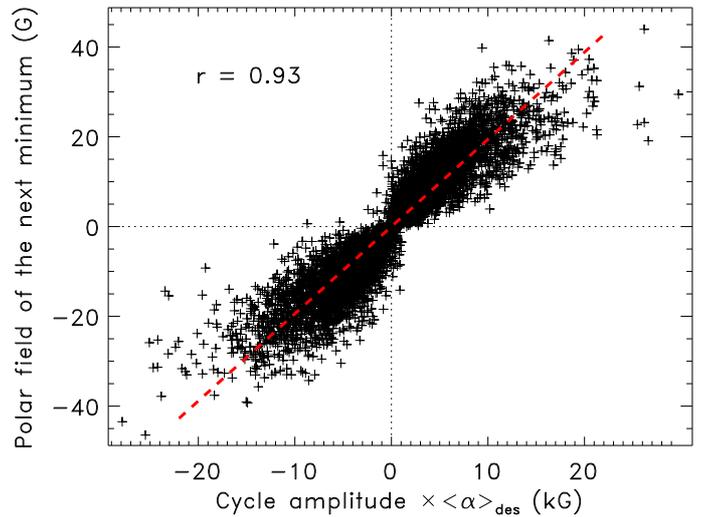}
   \caption{Correlation between the polar field at the cycle end and the
            dynamo-related parameter of the product of the amplitude of the toroidal field and descending phase averaged $\alpha$-effect. The dashed line shows the best linear fit.
              }
   \label{f6}
\end{figure}

Amplitudes of the magnetic cycles are largely controlled by polar fields of the cycle onset (Figs. \ref{f4} and \ref{f5}). These fields are produced in the course of preceding cycles. Figure~\ref{f6} shows a tight correlation between the polar fields $B_\mathrm{p}$ at the cycle end and the product $B_\mathrm{t}\langle \alpha \rangle_\mathrm{des}$ of the cycle amplitude $B_\mathrm{t}$ and the (normalised) time-averaged $\alpha$-parameter
\begin{equation}
    \langle\alpha\rangle = \frac{1}{T_2 - T_1}\int\limits_{T_1}^{T_2}
    \left( 1 + \sigma s(t)\right)\,\mathrm{d}t .
    \label{8}
\end{equation}
The low index in $\langle\alpha\rangle_\mathrm{des}$ means that $T_1$ and $T_2$ in Eq.\,(\ref{8}) are the times of maxima and commencements of modelled cycles, respectively. The dashed line in Fig.\,\ref{f6} shows the linear fit $B_\mathrm{p} = 1.94\,B_\mathrm{t}\langle \alpha \rangle_\mathrm{des}$. The correlation coefficient for the plot of Fig.\,\ref{f6} is high $r = 0.93$. The plot of Fig.\,\ref{f6} shows a slightly larger scatter if produced with the $\langle\alpha\rangle$ averaged over the entire cycle.
\subsection{High amplitude cycles}
The detection of numerous flares of very high energy
\cite[superflares of $\ga  10^{34}$\,erg;][]{Mea12,Sea13} among  solar-type targets of space telescope
{\sl Kepler} provoked a discussion of the possibility and hypothetical origin of exceptionally strong solar cycles \citep{Sea13PASJ,Cea14,KO16}. Figure\,\ref{f5} shows that cycles with magnetic energy about 20 times higher compared to the model without fluctuations were met in our simulations.

\begin{figure}
   \includegraphics[width=\hsize]{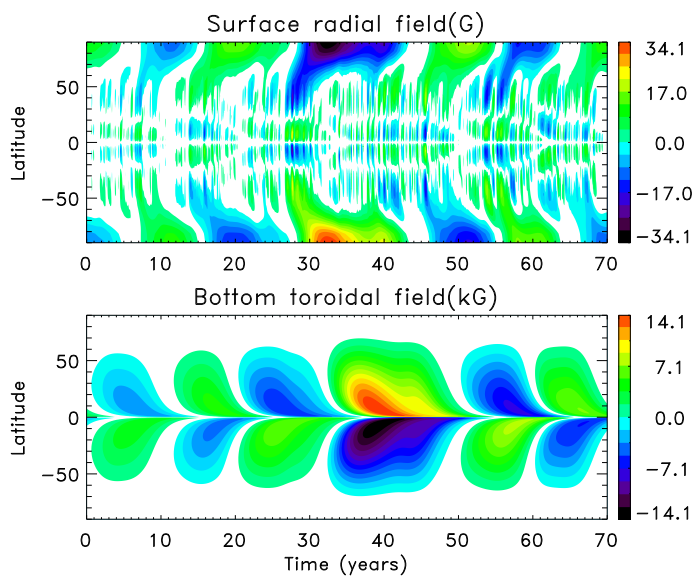}
   \caption{Time-latitude diagrams of the surface radial field ({\sl top panel})
            and the bottom toroidal field ({\sl bottom}) with an example of a strong cycle in the range between 30 years and 50 years.
              }
   \label{f7}
\end{figure}
\begin{figure}
   \centering
   \includegraphics[width=8 truecm]{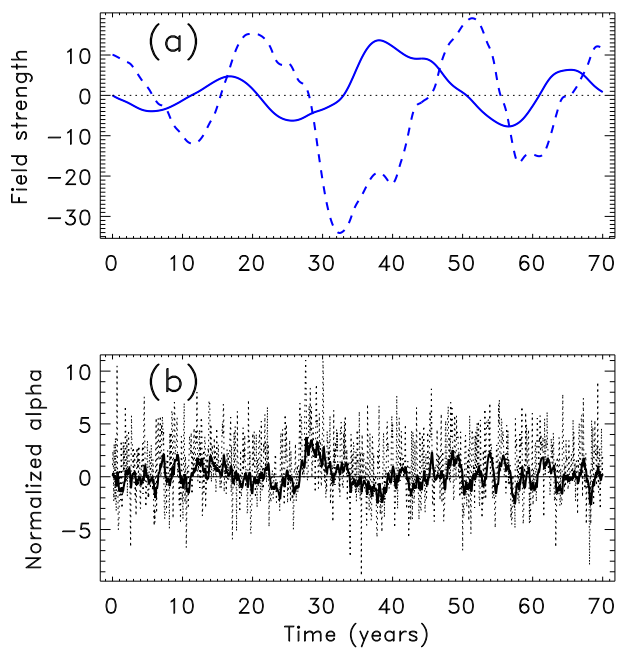}
   \caption{(a) Northern polar field (Gauss, dashed line) and toroidal
                field $B_\mathrm{t}$ (kilo-Gauss, full line). (b) Normalised alpha $1 + \sigma s(t)$ (dashed) and its annual running mean (full line), all for the same run as Fig.\,\ref{f7}.
              }
   \label{f8}
\end{figure}

Figures \ref{f7} and \ref{f8} give a characteristic example of such a strong cycle. The fine structure in the poloidal field of Fig.\,\ref{f7} at low latitudes is caused by fluctuations in the $\alpha$-effect. The fine structure is smoothed out by turbulent diffusion while the field is transported to the poles by the diffusion and meridional flow so that the polar field of Fig.\,\ref{f8} varies smoothly. Also the toroidal field of Figs.\,\ref{f7} and \ref{f8} is smooth. The origin of the strong cycle between 30 and 50 years in these figures can be seen from the plot of the fluctuating $\alpha$ in Fig.\,\ref{f8}b. The plot shows a strong positive fluctuation around the time of 30 years. This fluctuation produced an abnormally large poloidal field at the beginning of the strong cycle. The large poloidal field defined the large low bound for the toroidal field strength and energy of the coming cycle (Figs.\,\ref{f4} and \ref{f5}). The $\alpha$-fluctuation changed to a large negative at the beginning of the strong cycle thus delaying decay of the large poloidal field and increasing further the cycle energy.

This scenario is typical of the strong cycles of our simulations. The strong cycles are however not likely to cause superflares. The energy released in flares is by all probabilities related to the magnetic energy generated by internal solar dynamo. The functional form of the relation is not known however.
If we assume that the relation is linear and that the maximum energy $\sim 10^{32}$\,erg of solar flares ever observed corresponds to an  average cycle, then the maximum flare energy for the strongest cycles of our model is of the order $10^{33}$\,erg. This is even smaller energy than in earlier estimations by \citet{KO16}. Total dynamo-generated energy of Fig.\,\ref{f5} is much larger but the flux tubes raising from deep of the convection zone bring a minor part of the total energy to the surface active regions \citep{DC93}.

It can be noted that two high cycles between 20 and 50 years in Fig.\,\ref{f8}a show the asymmetry between their growth and decay phases discussed in Sect.\,\ref{S3.1}.
\subsection{Sudden drops in cycles amplitude}
The current 24th cycle of solar activity is much weaker compared to the preceding 23rd cycle. Similar or even much stronger drops in amplitudes between neighbouring cycles were met in our simulations. Figures \ref{f9} and \ref{f10} show a characteristic example.

\begin{figure}
   \includegraphics[width=\hsize]{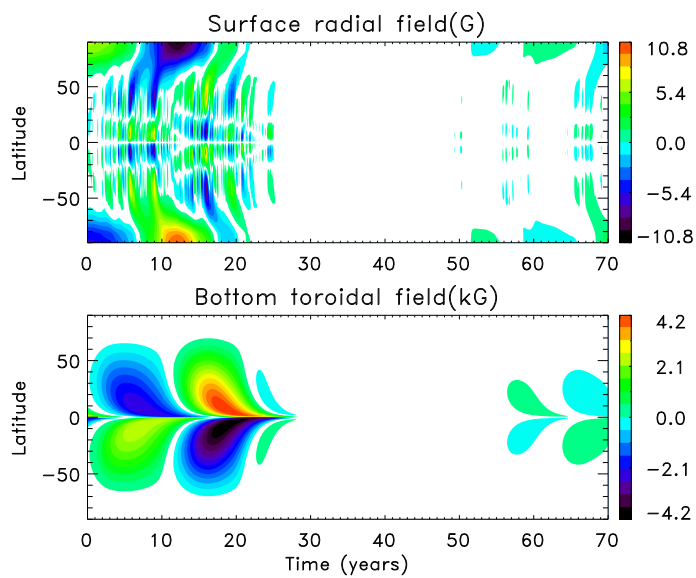}
   \caption{Time-latitude diagrams of the surface radial field ({\sl top panel})
            and the bottom toroidal field ({\sl bottom}) with an example of a
            sharp drop in cycle amplitude.
              }
   \label{f9}
\end{figure}
\begin{figure}
   \centering
   \includegraphics[width=8 truecm]{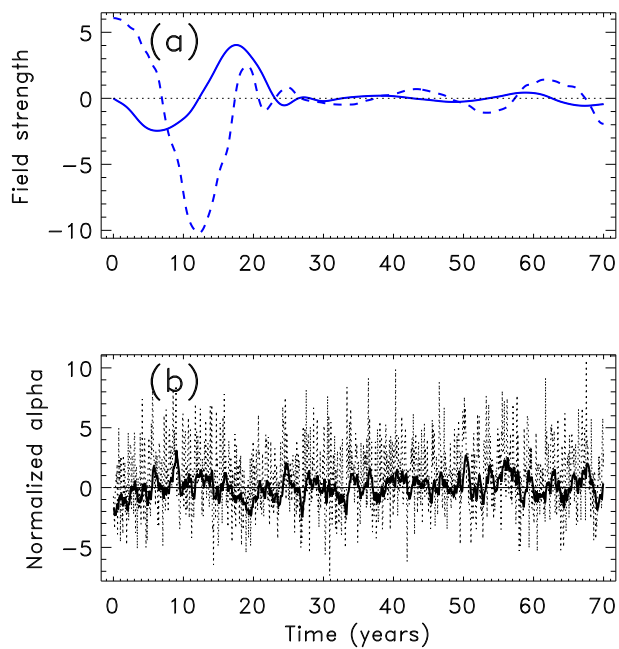}
   \caption{(a) Northern polar field (Gauss, dashed line) and toroidal
                field $B_\mathrm{t}$ (kilo-Gauss, full line).
                (b) Normalised alpha $1 + \sigma s(t)$ (dashed) and its annual
                running mean (full line), all for the same run as Fig.\,\ref{f9}.
              }
   \label{f10}
\end{figure}

A sharp drop followed the second of the full cycles shown in these figures. The drop was caused by negative fluctuation in the $\alpha$-parameter in the descending part of this cycle (Fig.\,\ref{f10}b). A usual reversal of the polar field occurred near to the cycle maximum. Growth of the reversed polar field was, however, hampered by the negative fluctuation. The fluctuation caused a surge of radial field of  old polarity to the poles seen in the top panel of Fig.\,\ref{f9}. This caused the  seed poloidal field for the next cycle to be small, braking the normal course of cyclic dynamo. The brake is manifested by two short  false cycles in the bottom toroidal field (Fig.\,\ref{f10}a). Magnetic cycles of normal duration recovered hereafter. Our dynamo model is slightly ($\simeq$10\%) supercritical. The recovery of normal cycle amplitudes in the simulated grand minimum of Figs. \ref{f9} and \ref{f10} was, therefore, slow.

The scenario of sharp onset and slow recovery was typical of grand minima in our simulations, although there were also cases of smooth onset. It is not certain whether the onset of the Maunder minimum in solar activity was sharp or smooth \citep{UMK00,Vea11}.
The statistics of grand minima and maxima of our simulations are similar to an earlier model by \citet{OCK13} and we do not re-discuss these statistics.
\subsection{Why is the current activity cycle low?}
The current 24th solar cycle is about 1.7 times weaker compared to the 23rd cycle in the sense of spot number (https://solarscience.msfc.nasa.gov/SunspotCycle.shtml). The reason for the low sunspot activity can be seen from Fig.\,\ref{f11} \citep[see][for data origin and processing used for this Figure]{MY14}. Polar fields of either hemisphere reversed near to the maximum of the solar cycle 23. Growth of the reversed field was however prohibited by poleward surges of fields of  old polarity that occurred in both hemispheres \citep{MGE15,GM17}. The surges originated from active regions with non-Joy tilts. As a result, polar fields at the end of the 23rd cycle were relatively small. The small poloidal field at the beginning of 24th solar cycle caused the weakness of the cycle. This interpretation agrees with \citet{JCS15}.

\begin{figure}
   \includegraphics[width=\hsize, height=7.9 truecm]{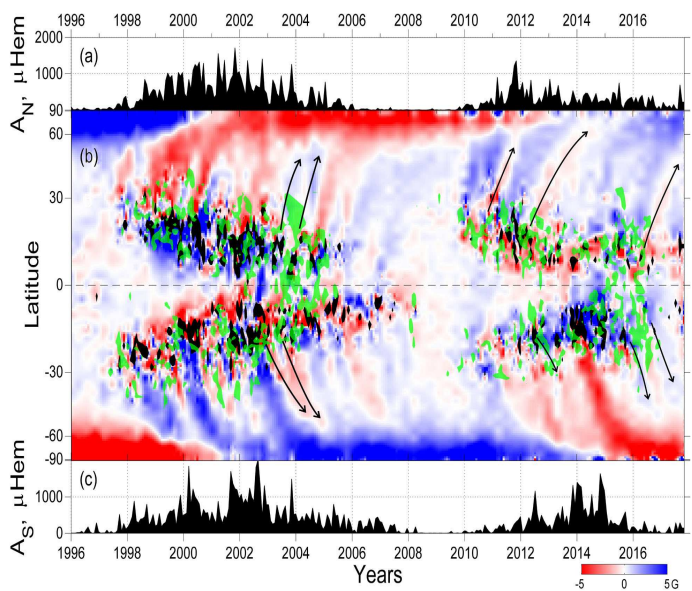}
   \caption{Sunspot areas for the northern (a) and southern (c) solar
            hemispheres, respectively, in the two latest activity cycles. Panel (b) shows the time-latitude diagram of the large-scale radial field. Domains of active regions with non-Joy tilts and regions of intense sunspot activity are overplotted in green and black, respectively. Arrows show the leading-polarity surges.
              }
   \label{f11}
\end{figure}

The non-Joy tilts correspond to a negative fluctuation in the BL mechanism. Observational Fig.\,\ref{f11} is to some extent similar to theoretical Fig.\,\ref{f9}, although with a less sharp decline in the observed activity between the neighbouring solar cycles.

Leading-polarity surges are also observed in the current cycle. Figure\,\ref{f11} shows the leading-polarity surges (marked with arrows) that alternate with the trailing-polarity surges. Such an alternation of opposite polarities suggests that the Sun’s polar field weakening continues in the current cycle.
\section{Conclusions}\label{Conclusions}
The correlation time of fluctuations in the BL mechanism for a generation of the poloidal magnetic field is close to the solar rotation period. This conclusion follows from a comparison of the statistics of computed dynamo-cycle periods with the distributions of periods of 36 directly observed and 119 reconstructed \citep{Nea15} solar cycles. Fluctuations in the BL-type $\alpha$-effect induce asymmetry in magnetic cycles of the dynamo model with the rise time of simulated cycles being on average shorter than the time of decline.

The effect of fluctuations on the amplitude of magnetic cycles depends on the phase of the cycle in which the fluctuations occur. Fluctuations in ascending phase increase the amplitudes of the toroidal field and magnetic energy. The increase is however moderate and the cycle amplitudes are mainly controlled by the strength of the poloidal (polar) field at the beginning of the cycle. The polar fields of the minima of the cycle define, therefore, low bounds on the amplitudes of the toroidal fields and magnetic energy of the following cycles (Figs.\,\ref{f4} and \ref{f5}).

The polar fields of the minima epochs are largely controlled by fluctuations in the BL mechanism in descending phase of preceding cycles. Positive fluctuations (increasing the $\alpha$-effect) increase the polar field and can lead to strong following cycles with tens of times larger magnetic energy compared to its representative value. The energy estimations however show that fluctuations in dynamo parameters cannot cause superflares of $\ga 10^{34}$\,erg on the Sun. \citet{Kea18} arrived at the same conclusion from different considerations. The origin of superflares on solar twins \citep{SKD00,Nea14} remains a puzzle.

Negative fluctuations in the descending phase of a cycle act to decrease the polar field at the cycle's end and the amplitude of the next cycle. This mechanism can produce sharp drops in amplitudes between neighbouring cycles. The slightly supercritical solar dynamo \citep{CS17,KN17MNRAS} recovers slowly from such drops (Fig.\,\ref{f10}). The drop in amplitude of the 24th solar cycle compared to the 23rd cycle can be explained by violations of Joy's law (negative fluctuations in the BL mechanism) by active regions in descending phase of the 23rd cycle.
\begin{acknowledgements}
The authors are thankful to an anonymous referee for pertinent and constructive comments. This work was supported by the Russian Foundation for Basic Research (projects  17-02-00016 and 17-52-80064) and by budgetary funding of Basic Research program II.16.
\end{acknowledgements}
\bibliographystyle{aa}
\bibliography{paper}
\end{document}